\documentclass[prl,twocolumn,twoside,showpacs,superscriptaddress]{revtex4}
\usepackage{graphicx,amssymb,amsmath,epsf}
\epsfclipon

\begin{document}

\title{Relevance of Metric-Free Interactions in Flocking Phenomena}
\author{Francesco Ginelli}
\affiliation{Service de Physique de l'\'Etat Condens\'e, CEA--Saclay,~91191~Gif-sur-Yvette,~France} 
\affiliation{Institut des Syst\`emes Complexes de Paris \^Ile-de-France, 57-59 rue Lhomond, 75005 Paris, France}%

\author{Hugues Chat\'e}
\affiliation{Service de Physique de l'\'Etat Condens\'e, CEA--Saclay,~91191~Gif-sur-Yvette,~France}

\date{\today}

\begin{abstract}
We show that the collective properties of self-propelled particles 
aligning with their ``topological'' (Voronoi) neighbors are 
{\it qualitatively} different from those of usual models where 
metric interaction ranges are used.
This relevance of metric-free interactions,
shown in a minimal setting, 
indicate that realistic models for the cohesive motion of cells, bird flocks,
and fish schools may have to incorporate them, 
as suggested by recent observations.
\end{abstract}

\maketitle

Active matter, where energy is spent locally to produce motion, such
as in the cytoskeleton of living cells, displays a number of fascinating
phenomena and raises important new theoretical questions \cite{SRIRAM-REVIEW}.
One such general problem is that of the emergence of collective motion in large groups
of self-propelled objects. 
Alongside the more traditional studies of animal behavior, the
physics community focused, 
not surprisingly, on minimal models and their 
universal properties, following the seminal work of Vicsek {\it et al.}
 \cite{VICSEK}.
In many of these works, 
models consist of interacting particles 
and rely on metric thresholds and cut-offs to delimit various
interaction zones \cite{PVG2002}.

Recent facts, however, point towards a metric-free, topological
determination of the ``neighbors'' with whom a given individual interacts.
The StarFlag project provided snapshots of reconstructed three-dimensional 
flocks of starlings of up to a few thousand birds. Statistical analysis
revealed that a typical starling significantly interacts 
with its 7 or 8 closest neighbors,
located at rather well-defined angular positions, but irrespective of their
actual distance \cite{STARFLAG}. The authors then concluded to a ``topological''
determination of neighbors, although their results, obtained on flocks
with a rather homogeneous density, can also be explained by a mere 
global rescaling of distances by each flock's density.
Another hint at the relevance of metric-free interactions can be found in the 
recent work on crowd dynamics by Moussa\"{\i}d {\it et al.} \cite{MEHDI}
who argue that pedestrians make decisions
based on the ``angular landscape'' formed by surrounding others,
who may screen out otherwise close neighbors.
Finally, such a screening effect is probably important in
(moving) cohesive groups of cells with direct cell-to-cell interactions.

In this Letter, we investigate the influence of metric-free, ``topological''
determination of neighbors on collective properties of self-propelled
particles in a minimal setting. 
We introduce a ``topological Vicsek model'' in which constant-speed 
particles align their velocity with that of others forming 
their first Voronoi shell. 
We show that the introduction of Voronoi neighbors changes qualitatively
the collective properties of the original model. 
At finite density, true long-range order arises but,
contrary to the metric model,
no density segregation is present and
the transition to collective motion is continuous
with non mean-field critical exponents that we determine numerically 
in two dimensions.
Large flocks evolving in an infinite domain move and spread diffusively,
but quasi-long-range order nevertheless arises.

In the original, ``metric'', Vicsek model point particles align locally, 
in competition with some noise. 
The interaction consists in taking the
(polar, ferromagnetic) average of velocities over all particles within
unit distance. Two types of noise have been studied: 
the ``angular'' noise consists in adding a random angle to the orientation
of the averaged velocity; with the ``vectorial'' noise, 
one adds a random vector
to the average velocity before taking the orientation of the resulting vector
to stream particles (for details, see, {\it e.g.} \cite{CHATE}).
In all cases, particles perform uncorrelated random walks at strong noise, 
while they align perfectly in the absence of noise. 
An order-disorder transition happens in between, 
which was originally found continuous
by Vicsek {\it et al.} \cite{VICSEK}.
If the existence, even in two dimensions, of a fluctuating 
collective motion phase with long-range order and algebraic correlations
was established early by Tu and Toner \cite{TT}, 
later studies \cite{CHATE,BGD,IHLE} revealed that 
it is subdivided in two regions and that the transition is discontinuous
due to the spontaneous emergence of dense, ordered, traveling 
solitary bands or sheets.
The importance of these structures has been recognized implicitly 
by a number of groups \cite{FRIENDS}
which tried to frustrate their emergence 
 because they question
the discontinuous character of the transition in the angular noise case.
In the following, we restrict ourselves to the case of 
vectorial noise for which the discontinuous nature of the transition 
is admitted. 

The topological Vicsek model with vectorial noise is defined like
its metric counterpart:
$N$ point particles move off-lattice at constant speed $v_0$. 
In two dimensions (to which we restrict ourselves below), 
particle $j$ is defined by its 
(complex) position $\mathbf{r}_j^t$ and 
orientation $\theta_j^t$, updated at unit time steps according to 
\begin{eqnarray} 
\label{motion_angle} 
\theta_j^{t+1}&=& 
\arg \left[\sum_{k\sim j} \exp{i\theta_k^t} + \eta {\cal N}^t_j  {\bf\xi}_{j}^{t} \right] \\ 
\mathbf{r}_j^{t+1}&=& \mathbf{r}_j^{t}+v_0\, \exp{i\theta_k^{t+1}} \, , 
\label{motion_pos} 
\end{eqnarray} 
where $\xi$ is a random unit vector and $\eta$ the noise strength.

\begin{figure}[tp]
\includegraphics[width=8.6cm,clip]{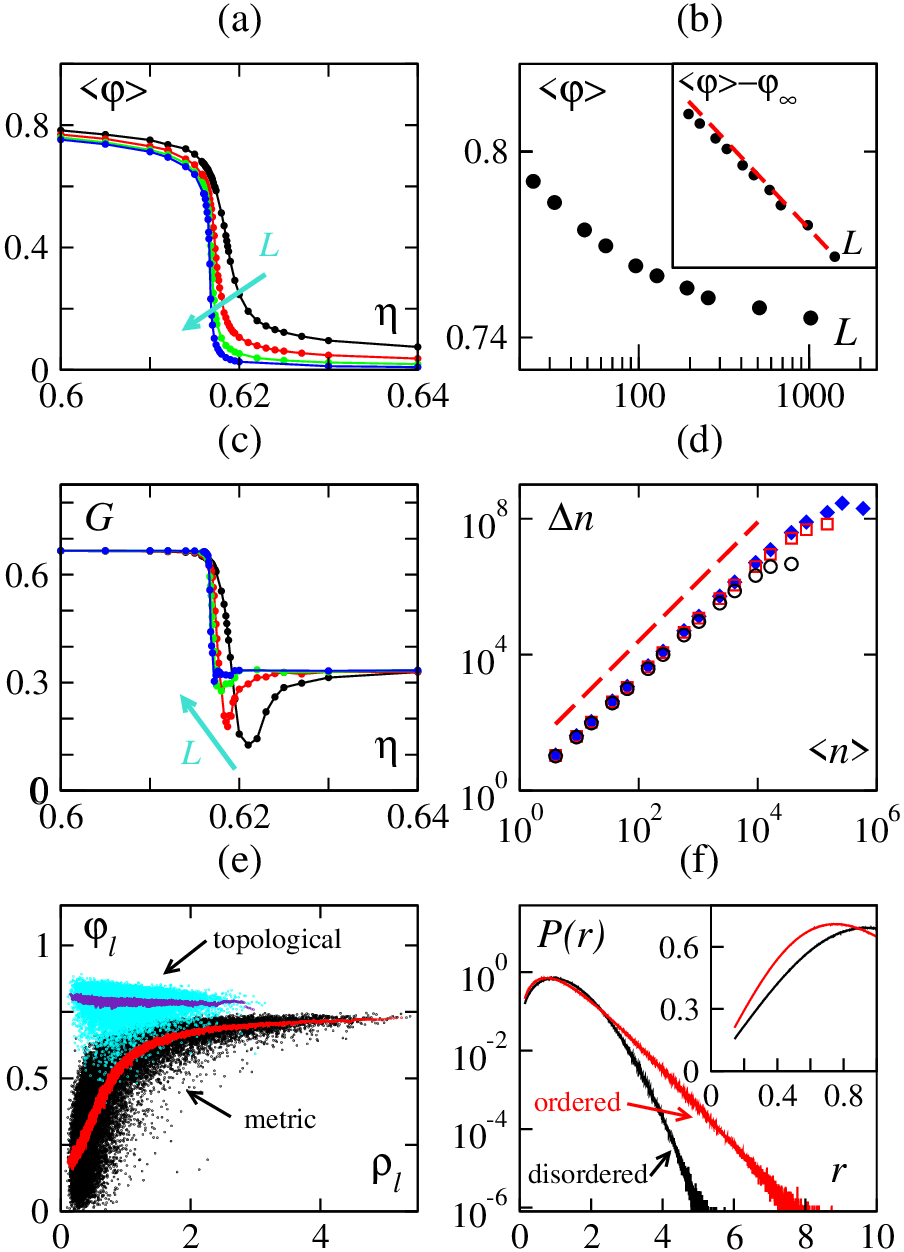}
\caption{(color online)
(a) time-averaged order parameter $\langle\varphi\rangle$ vs $\eta$
for different system sizes $L$.
From top to bottom 
(the cyan arrow points from smaller to larger systes sizes): 
$L=32$, $64$, $128$, $256$.
(b) $\langle\varphi\rangle$ vs $L$ at $\eta=0.6$ (logarithmic scales).
Inset: same data from which $\varphi_{\infty}=0.741166$ has been subtracted.
The dashed line marks a power law decay as $L^{-2/3}$.
(c) Binder cumulant $G$ vs noise amplitude $\eta$
for different system sizes $L$ (same symbols and colors as in (a)).
(d) Giant density fluctuations in the ordered phase.
$\Delta n$ vs $n$ at $\eta=0.6$ for different system sizes:
$L=256$ (circles), $512$ (squares), $1024$ (blue).
The dashed red line has a slope $1.75$. 
(e) Scatter plot of local order parameter $\varphi_l$ versus
local density $\rho_l$ computed in boxes of size $\ell = 16$ in
the ordered phase. Black dots: Vicsek metric model with
interaction distance $r_0=1$ and parameters $v_0=0.5$, $\eta=0.6$,
$\rho=2$ and $L=256$. Cyan dots: topological model with $v_0=0.5$, $\eta=0.6$,
$\rho=1$ and $L=512$. The lines passing through the middle of 
these two clouds of points are local averages in a moving window in
$\rho_l$.    
(f) (normalized) distribution $P(r)$ of distances to
Voronoi neighbours ($L=128$). 
Black line: $\eta=0.8$ (disordered); red line $\eta=0.55$ (ordered).
Both cases yield typical distances of order $r=1$.
Inset: same in linear scales, showing that $P(r) \sim r$ as $r \to 0$.
}
\label{fig1}
\end{figure}
The key difference with the metric case is that the neighbors $k$ 
are chosen to be the ${\cal N}^t_j$ particles forming the first shell around particle $j$
in the Voronoi tesselation constructed from the particle positions at time $t$.
With respect to its metric counterpart, this model has one less parameter 
since there is no interaction range. 
In most of the following, we used square domains of linear size $L$ 
with periodic boundary conditions.
Without loss of generality, we rescale lengthscales
in order to fix the density $\rho=N/L^2=1$. We are then left with two
parameters, the microscopic speed $v_0$, which we keep constant
at $v_0=0.5<1/\sqrt{\rho}=1$, and our main control
parameter, the noise strength $\eta$.

At large $\eta$, noise dominates the interaction, and one observes essentially
random configurations. Decreasing the noise strength, 
the order parameter $\varphi(t)=|\langle\exp i\theta^t_j\rangle_j|$ 
increases to order one values  (Fig.~\ref{fig1}a).
For a fixed noise strength in the ordered side, $\varphi$ decreases slightly
with $L$, but slower than a power-law (Fig.~\ref{fig1}b): our topological model,
like the original Vicsek model, gives rise to true long-range polar order.
Its ordered phase, though, is qualitatively different from that of the
metric model: here, there is no coupling between local density and
local order  (Fig.~\ref{fig1}e), because neighbors in a sparse region
are never disconnected, and thus low density does not necessarily
induces disorder. The segregation mechanism which allows
eventually for the emergence of bands in the metric model is absent. Indeed,
scanning sizes up to $L=1024$, no band-like structure was detected.

Giant number fluctuations ---a signature of fluctuating ordered
active phases \cite{TTR-REVIEW}--- 
are nevertheless present, as in metric models:
the variance $\Delta n$ of the number of particles
present in square boxes of linear size $\ell$ scales like $n^\alpha$ with 
$\alpha\simeq 1.75 > 1$ where $n=\rho\ell^2$ is the average number of 
particles in the box (Fig.~\ref{fig1}d). This value of $\alpha$ is,
however, significantly different
from that known for the metric polar models,
$\alpha=\frac{8}{5}$ \cite{CHATE,TONER}. 
The exact origin of this discrepancy, beyond the above remark about
the absence of a basic segregation mechanism, remains unclear: 
a good reason would be the occurence of long-range connections, 
allowed by the topological determination of neighbors, but 
this is not the case: 
distances to (Voronoi) neighbors are exponentially distributed
in the ordered phase, which is qualitatively different from
the Gaussian tail observed in the disordered phase,
but both yield a small typical interaction range (Fig.~\ref{fig1}f).

\begin{figure}[tp]
\includegraphics[width=8.4cm,clip]{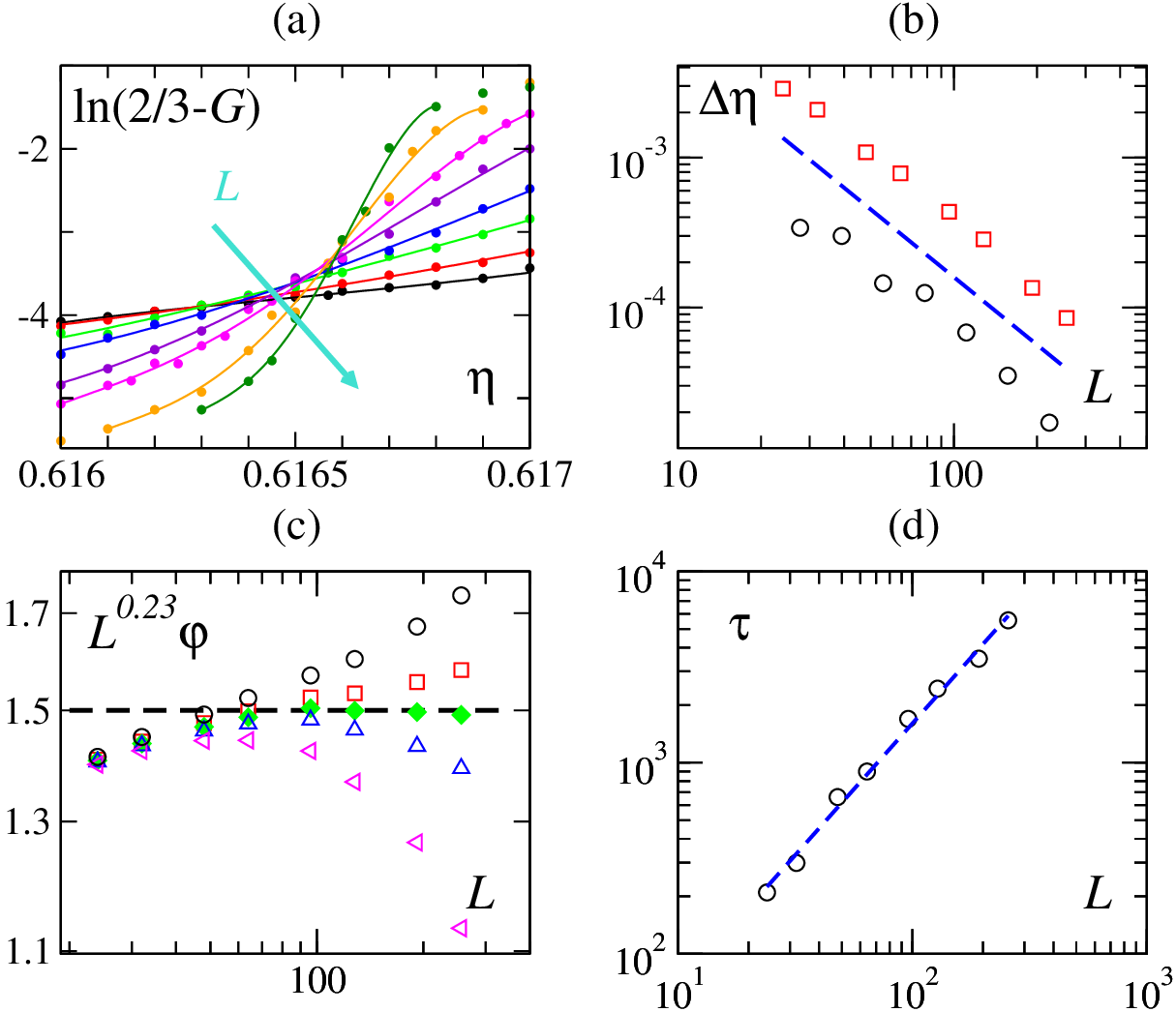}
\caption{(color online)
(a) Zoom on the crossing area of the $G(\eta)$ curves obtained at different 
$L$ values ($L=24,32,48,64,96,128,192,256$).
(The variable $\log(\frac{2}{3}-G)$ was chosen to produce the 
rather good quality polynomial fits shown with solid lines).
(b) $\Delta \eta = |\eta_{\rm c}(L)-\eta_c^\infty|$ vs. $L$
for $\eta_c^\infty=0.61661$.
Circles: $\eta_{\rm c}(L_{12})$
is the crossing point between the $G(\eta)$ curves of panel (a) 
for two consecutive sizes $L_1$ and $L_2$.
Squares: $\eta(L)$ is given by the position of the maximum
of the susceptibility.
The dashed line marks a power law decay as $L^{-1.5}$.
(c) $\varphi \times L^{0.23}$ vs $L$ for different $\eta$ values 
 using fitted order parameter curves
(from top to bottom $\eta=0.6165$, $0.61658$, $0.61661$,
$0.61664$, $0.6171$). The flat dashed black line is a guide for the eye.
(d) Autocorrelation time vs. $L$ at the critical point. 
The dashed line has a slope of $1.38$. }
\label{fig2}
\end{figure}

While they are no proof, the above findings, and in particular
the absence of bands, hint that the transition
of our topological model might not have the discontinuous character observed
in its metric counterpart. 
The behavior of the so-called Binder cumulant 
$G=1-\langle\varphi^4\rangle_t/(3\langle\varphi^2\rangle_t^2)$ 
confirms this observation:
when decreasing the noise strength $\eta$, $G$ exhibits
a dip towards values smaller than $\frac{1}{3}$ (the expected value in the 
disordered phase) before reaching  $\frac{2}{3}$, the ordered value
(Fig.~\ref{fig1}c).
If this dip were to sharpen and deepen to negative values 
when increasing system sizes, it would be the signature of a discontinuous
transition \cite{Binder1997}. 
But here, on the contrary, it gradually disappears
as $L$ is increased (Fig.~\ref{fig1}c), definitely ruling out this possibility:
in the infinite-size limit, the Binder cumulant varies monotonically
through the transition, which is thus continuous.
To estimate the values of the associated critical exponents, we performed 
a classic finite-size-scaling study \cite{FSS}, 
measuring the first moments of 
the order parameter as functions of $\eta$ for various systems sizes. 
The correlation time $\tau(\eta,L)$, estimated from the exponential decay
of the autocorrelation function of the order parameter, was measured 
systematically and we made sure that our runs were always longer than
 $10^3 \tau$.
To locate the critical point, we determined the $\eta$ values at which
the $G(\eta)$ curves obtained for different system sizes $L$ cross each other
(Fig.~\ref{fig2}a). 
Without corrections to scaling, these curves should cross at the asymptotic
critical point $\eta_{\rm c}^\infty$ where the Binder cumulant takes a universal
value $G_{\rm c}^\infty$. Here, the crossing points $\eta_{\rm c}(L_1,L_2)$
estimated for two consecutive sizes
$L_1$ and $L_2$ are not identical but quickly converge, 
as $L_{12}\equiv\sqrt{L_1 L_2}$ increases, to $\eta_{\rm c}^\infty=0.6166(1)$. 
If due to ``ordinary''
corrections to scaling, the convergence of the crossing points
should be described by 
$\eta_{\rm c}(L_1,L_2)=\eta_{\rm c}^\infty+a L_{12}^{1/\nu}$ and that of $G_{\rm c}(L_1,L_2)$ 
should be
$G_{\rm c}(L_1,L_2)=G_{\rm c}^\infty+b L^{-\omega}$ where $\omega$ is the first correction 
to scaling exponent \cite{FSS}. 
These scaling laws are well verified by our data using
the above estimate of $\eta_{\rm c}^\infty$ (Fig.~\ref{fig2}b, not shown
for $\omega$). Similarly,
the location of the maximum of the susceptibility 
$\chi=L^d(\langle\varphi^2\rangle_t-\langle\varphi\rangle_t^2)$ converges to
$\eta_{\rm c}^\infty$ with the exponent $1/\nu$, 
but here from above whereas it is 
from below for the effective thresholds $\eta_{\rm c}(L_1,L_2)$
determined using the crossings of the Binder cumulant curves.
Using these two approaches to $\eta_{\rm c}^\infty$ 
allows us to narrow our estimate
to $\eta_{\rm c}^\infty=0.61661(3)$ and yields $1/\nu=1.5(1)$ 
(Fig.~\ref{fig2}b).
Next, we verified that our finite-size estimates of the order parameter at the 
critical point satisfy $\varphi(L,\eta_{\rm c}^\infty) \propto L^{-\beta/\nu}$
(Fig.~\ref{fig2}c), yielding $\beta/\nu=0.23(3)$.
Making use of the usual so-called hyperscaling relation $2\beta-\gamma=\nu d$
where $d$ is the space dimension, the above results allow us to complete 
our estimates of the static exponents yielding $\gamma/\nu=1.54(6)$.
Additional consistency checks of the above numbers (using various
other quantities) were performed. For instance, 
the scaling of the maximum of the susceptibility $\chi_{\rm max}(L)\propto L^{\gamma/\nu}$
is rather nicely verified, yielding an estimate of $\gamma/\nu=1.49(5)$ 
compatible with the above numbers (not shown). 
Finally, the divergence with $L$ of the correlation time 
estimated at criticality, $\tau(\eta_c^\infty,L) \propto L^z$,
yields an estimate of the dynamical exponent $z=1.38(5)$ (Fig.~\ref{fig2}d).
A summary of our estimates is in Table~\ref{table}. 
They indicate that the critical properties of
our model do {\it not} correspond to mean-field values (where, for instance
$\beta=\frac{1}{2}$), nor to any known universality class.


\begin{table}[tp]
\caption{\label{table} Critical exponent and Binder cumulant 
values.}
\begin{ruledtabular}
\begin{tabular}{cccccc}
$1/\nu$   & $\beta/\nu$ & $\gamma/\nu$ & z &  $\omega$ & $G_{\rm c}^\infty$  \\ \hline
1.5(1) &  0.23(3)    &    1.49(5)   & 1.38(5) & 0.67(20) & 0.615(2)
\end{tabular}
\end{ruledtabular}
\end{table}

\begin{figure}
\includegraphics[width=8.6cm,clip]{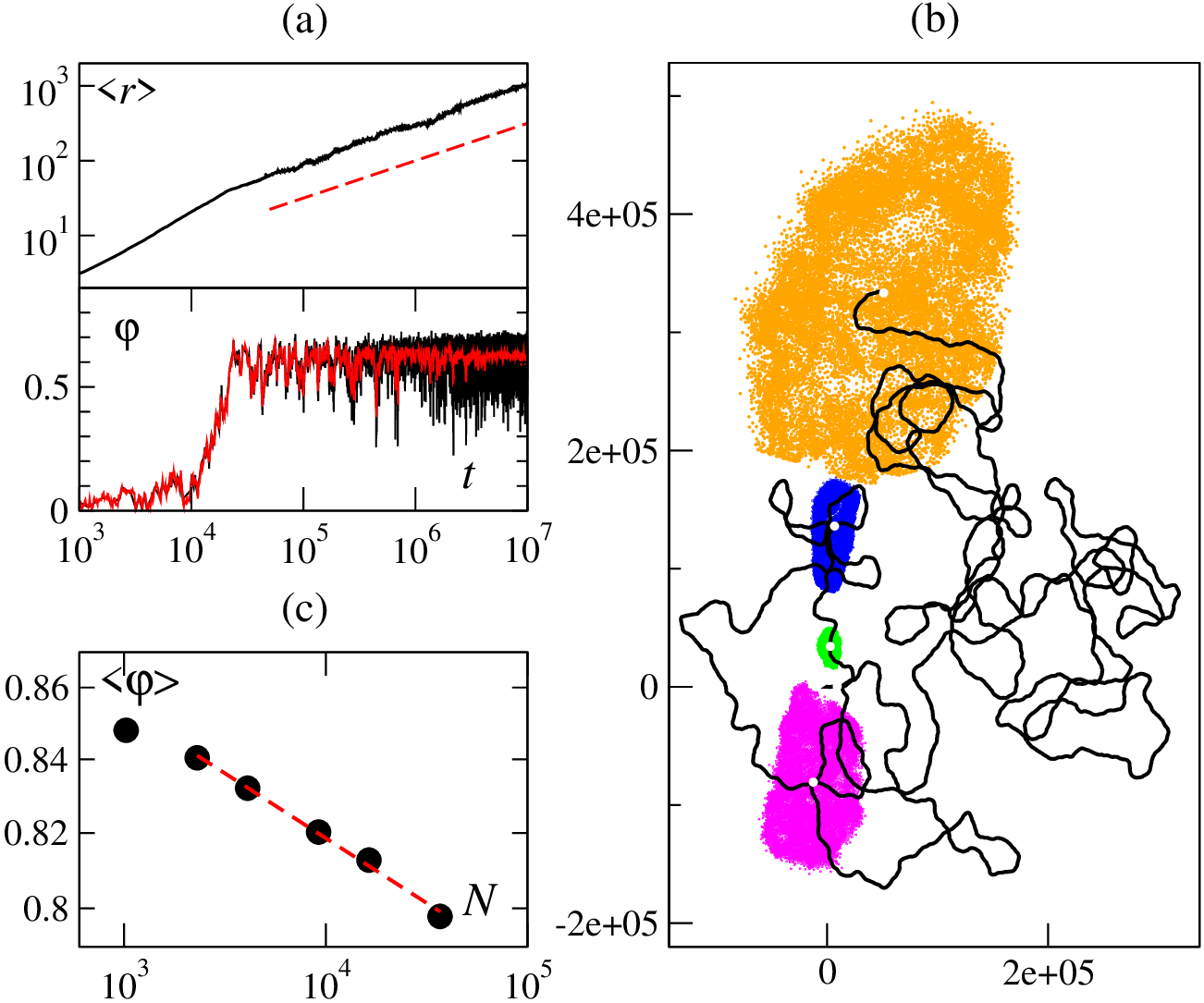}
\caption{(color online) Ordered flock in an infinite domain 
(initial conditions: random positions and random orientations in a unit circle).
(a) mean distance $\langle r \rangle$ between Voronoi neighbors (top)
and $\varphi(t)$ (bottom) vs. time 
($N=2^{14}$, $\eta=0.61$): after transient, the flock is ordered and 
spreads diffusively ($\langle r \rangle\sim \sqrt{t}$), indicated by the
dashed line). The red data in the bottom panel is a moving average on a window
of fixed length in $\ln(t)$. 
(b) Trajectory of the center of mass for $t\in [1,10^7]$ plus snapshots of 
the flock in the asymptotic regime at $t=32\times 10^3$ (red), 
$128\times 10^3$ (green),
$512\times 10^3$ (blue) $2048\times 10^3$ (magenta) and $10^7$ (orange) 
(same run as in (a)). The trajectory shows no sign of ``aging'' as the flock
expands diffusively.
(c) $\langle\varphi\rangle$ vs $N$ in logarithmic scales 
($\eta = 0.5$, time averages over $10^7$
timesteps after discarding an transient of $10^6$). 
Dashed line: power law with exponent $-0.074$.
}
\label{fig3}
\end{figure}

The transition to collective motion in the Vicsek topological model
and its ordered phase at finite density are thus
different from that of its metric counterpart.
Our model also possesses non-trivial properties in the zero-density
limit of a large group of $N$ particles evolving in an infinite domain.
Even though no attractive interaction is present, subgroups 
can never detach to become independent, because of the topological
determination of neighbors. This insures a weak cohesion:
a finite group typically expands diffusively at large times, 
but is nevertheless able to order at low enough noise strengths 
(Fig.~\ref{fig3}a),
something obviously impossible in metric models without 
attractive interactions.
Remarkably, the order parameter remains statistically stationary,
even on very long time scales, in spite of the constant 
expansion of the flock. 
The center of mass of the particles performs a persistent random walk
whose turns mostly reflect the fluctuations of the order
parameter and are not accompanied by large-scale changes in the group
structure or shape (Fig.~\ref{fig3}b).
In other words, the orientation axes of the particles turn, but there
is hardly any rotation of mass, just marginal changes in the
shape and the Voronoi tesselation. 

Thus, if space is rescaled by a dynamical length reflecting the
expansion of the group ($\ell\sim\sqrt{t}$), 
our system is mapped onto one where the microscopic velocity
now goes to zero like $1/\sqrt{t}$, evolving
 on a (quasi-) static Voronoi tessellation 
network: up to the rare rearrangements mentioned above,
this is an XY model.
Indeed, as for the two-dimensional XY model, only quasi-long-range order
arises in the zero-density limit of our topological model:
Increasing $N$, the average order parameter decreases slightly,
as in the finite-density case, but this time goes
to zero as a power-law (Fig.~\ref{fig3}c).

To summarize, we have introduced, motivated by recent observations
on animal group motion, a ``topological'' Vicsek model in which self-propelled
particles align with their Voronoi neighbors. 
We obtained similar results for the model
in which particles interact with their $n$ (metrically) 
closest neighbors  \cite{TBP}.
But this later case relies on distances and involves the 
additional parameter $n$, so we focused on the model presented here.
At finite density, these ``topological'' flocking models
 exhibit collective
properties qualitatively different from those of ``metric'' ones: the 
ordered moving phase shows novel long-range correlations (Fig.~\ref{fig1}d),
even though no direct long-range connections emerge (Fig.~\ref{fig1}f),
and the transition to collective motion exhibits critical properties 
different from known universality classes.

Beyond their importance in the general context 
of non-equilibrium order-disorder transitions, our results
indicate the relevance of ``topological'' interactions at the collective
level, following recent observations of their existence locally.
Since Voronoi neighbors remain at relatively well-defined distance
from each other,
our approach could also provide a novel framework for modeling cohesive
assemblies of cells, taking the Voronoi polygon around each particle
as the effective extension of a ``cell''.

Nevertheless, realistic ``topological'' models of bird flocks, 
fish schools, or cell assemblies must incorporate cohesive interactions
(e.g. pairwise attraction/repulsion) in order to account for the dynamics
of large but finite groups in open domains. We are currently pursuing
this line of research, together with theoretical investigations about
the novel scaling laws uncovered here.

We acknowledge the use of the CGAL package for constructing
Voronoi tesselations. Work funded in part by the French ANR projects
Panurge and DyCoAct.


\begin{thebibliography}{50}
\bibitem{TONER} The effective $\alpha$ values
recorded at fixed $L$ are larger than $\frac{8}{5}$ and
converge, as $L\to\infty$, 
to the asymptotic value (estimated at $1.75$) {\it from below}. Hence
one can be sure this value is different from $\frac{8}{5}$.
Note also that the metric model value $\alpha=\frac{8}{5}$ 
is {\it not} explicitly calculated in
\cite{TT}. It can nevertheless be simply deducted from these calculations. 
J. Toner and S. Ramaswamy, private communication.
\end{thebibliography}
\end{document}